				\documentstyle[12pt,aaspp4]{article}

\begin{document}
\slugcomment{To appear in the Astrophysical Journal, January 1998}
				\title{
An X-Ray Temperature Map of the Merging Cluster Abell~2255
				}
				\author{
David S. Davis			}
				\affil{
Center for Space Research, MIT, MS 37-662B, Cambridge, MA 02139
				}

\and
				\author{
Raymond E. White III     
				}

                                 \affil{
Department of Physics \& Astronomy, University of Alabama, Tuscaloosa, 
AL 35487-0324   
				}

                                \begin{abstract}
We present a spatially resolved map of intracluster gas temperatures
in the rich cluster Abell~2255, using $ROSAT$ PSPC data. 
The intracluster gas is strongly non-isothermal and the temperature
distribution supports the conclusions of Burns et al. (1995) that Abell~2255
is currently undergoing a merger. The hottest regions are near the core,
northwest of the center. A map of the ratio of softer 
to harder X-ray photons reveals an anomalously soft region northeast
of the center, coincident with a radio ridge discovered by Burns et al (1995).
We suggest that this cooler region is associated with an infalling 
group of galaxies which is giving rise to the radio ridge.
Meanwhile, a much more substantial recent merger between comparably sized 
subcluster components can account for the more global merger signatures,
which include the extensive central radio halo,
the large scale gas temperature variations,
the pair of central dominant ellipticals, 
elliptical X-ray isophotes offset from these dominant galaxies, and 
the very high velocity dispersion of galaxies in the cluster.

				\end{abstract}

				\clearpage

				\section{
Introduction
				}
Clues to the evolution of galaxy clusters can be found in the detailed 
dynamics of the hot gas and galaxies they contain. In hierarchical models for 
large-scale structure formation, clusters evolve by accreting 
small groups of galaxies and other clusters, a process that continues
at the present epoch.
X-ray signatures of a cluster merger include:
an X-ray peak offset from the peak of the galaxy distribution;
an elongated X-ray core; 
non-isothermal and asymmetric temperature distributions;
and the absence of a central cooling flow.
These merger signs can be identified using $ROSAT$ PSPC and $ASCA$ data, 
which provide spatially resolved X-ray imaging and spectroscopy. 
Clusters thought to have X-ray evidence of a recent (or ongoing) merger 
include Abell~2256 (Briel et al.  1991), Abell~754 (Henriksen \& 
Markevitch 1996; Henry \& Briel 1995; Zabludoff \& Zaritzky 1995), and Coma 
(White, Briel \& Henry 1993). X-ray temperature maps for Abell~2256 and 
Abell~754 show complex temperature structure,  most naturally explained 
if these clusters are undergoing a merger.

Abell~2255 is a rich cluster which shows several signs that it is 
undergoing a merger event. Some of the first evidence for a merger came from 
early $Einstein$ X-ray data: Jones \& Forman (1984) found that Abell~2255 has 
one of the largest X-ray core radii in their cluster sample. 
Stewart et al. (1984) further analyzed $Einstein$ IPC data and
concluded that Abell~2255 does not contain a cooling flow, 
which is unusual for a such a rich cluster (Edge, Stewart \& Fabian 1992). 
It is thought that a major merger would disrupt a pre-existing cooling 
flow and increase the core radius (Roettiger, Burns \& Loken 1993).
More recently, Burns et al. (1995) found from {\it ROSAT All Sky Survey}
($RASS$) data that the X-ray peak of Abell~2255 is not centered
on the brightest cluster galaxies but is offset by $\sim2^\prime$.
Thus, the X-ray data indicate that this cluster has recently undergone 
or is currently undergoing a merger. 
Feretti et al (1997) recently analyzed the $ROSAT$ PSPC pointed observation
of Abell~2255 and came to similar conclusions.

Abell~2255 has a very large velocity dispersion, $\approx$ 1200 km 
s$^{-1}$, which may indicate that it is dynamically unrelaxed, despite the 
lack of obvious spatial or kinematical substructure
(Stauffer et al. 1979).  Abell~2255 contains
two comparably bright central dominant galaxies, which is reminiscent
of the Coma cluster.  Having a pair of central dominant galaxies
is thought to be a sign that the cluster is the 
merger product of two similarly sized subclusters, each of which
had its own central dominant galaxy before the merger (Davis \& Mushotzky 
1993; Bird 1994).

Abell~2255 is also one of the rare clusters which contains a radio halo 
(Jaffe \& Rudnick 1979; Hanisch 1982; Feretti, et al 1997). 
The most recent radio observations (Feretti, et al 1997) show that it is 
$\sim1 h_{50}$ Mpc in diameter and unpolarized (which distinguishes
it from the numerous polarized tailed sources in this cluster).
Accounting for the rarity of cluster radio halos, as well as their varied 
morphologies and their energy source(s) have been long-standing puzzles. 
One of the main problems has been understanding how the relativistic electrons
which power radio halos get into the outer regions of the clusters 
where radio halos tend to extend. 
Tribble (1993) recently proposed that cluster mergers can provide 
the energy source for accelerating electrons throughout clusters.
Given the short lifetimes of relativistic electrons, it is 
thought that radio halos are indicative of very recent or $ongoing$ cluster 
mergers.

We present $ROSAT$ PSPC results which strengthen the case for a 
merger in Abell~2255. The PSPC data show that the cluster has a
complex temperature structure, which indicates that the cluster
is not in hydrostatic equilibrium. This, along with the gas and galaxy
peaks being offset, the elongated X-ray isophotes in the inner
region, and the existence of a radio halo, suggests that Abell~2255 is 
currently experiencing a merger. 
We describe the PSPC data in \S2 and \S3 and discuss its analysis in \S4. 
In \S5 we discuss the implications of the data, review the merger
evidence in this cluster, and summarize our conclusions.

				\section{
X-Ray Data                
				}
Burns et al (1995) previously analyzed the 3.5 ksec of {\it ROSAT All 
Sky Survey} PSPC data for Abell~2255. We analyze instead 
the $ROSAT$ PSPC's 14.5 ksec pointed observation of Abell~2255.
We corrected the X-ray imaging data for spatial variations in the 
exposure and vignetting, using the procedures outlined in Snowden et al (1994).
The PSPC data were selected from observation times when the 
{\it Master Veto Rate} ($MVR$) was within recommended bounds: $20<MVR<170$.  
The spectra for each of the regions discussed below was extracted from the
filtered data. 
The background for each spectrum was obtained from an annulus around 
the cluster extending from 30\arcmin$\,$ to 40\arcmin$\,$ from its center.
All detectable point sources were excluded from both source and background 
spectra, using a region size appropriate for the off-axis angle of the source. 
Source and background spectra were also corrected for vignetting 
effects and for the residual particle background (in the manner prescribed 
by Snowden et al. 1994), the latter of which is proportional to the $MVR$. 
Finally, the extracted spectra were rebinned so that each channel has a 
minimum of 25 counts. We also investigated the effects of using 
background spectra extracted from regions further from the cluster center, 
which reduces the cluster's contamination of the background, but increases the 
vignetting correction; no significant difference was found in the results. 

				\section{
X-ray Morphology
				}
The morphology of the X-ray gas in Abell~2255 can be seen in Figure 1,
which shows the PSPC image lightly smoothed with a $\sigma=30^{\prime\prime}$
Gaussian. 
The X-ray contours are elongated in an east-west direction in the center 
and appear to be more circular in the outer regions. 
The X-ray peak of this pointed observation is slightly offset from the 
brightest cluster galaxy. 
Fitting the X-ray surface brightness of the cluster confirms that the 
isophotes are offset from the central dominant galaxies and are fairly flat, 
with an ellipticity of $\sim$0.5. At larger radii from the center
($\sim$ 2--15\arcmin$\,$), we find the X-ray centroid is stable, but offset 
from the brightest cluster galaxy by a larger amount, $\sim$1\arcmin$\,$. 
The position angle of the X-ray isophotes are predominately 
east-west and the isophotal ellipticities are $>$0.25 inside 2\arcmin$\,$. 
Burns et al. (1995) report that the peak of the X-ray $RASS$ 
image is displaced from the brightest cluster galaxy by about
2\arcmin$\,$. We do not find such a large offset in the pointed observations.

				\section{
Spectral Analysis
				}
We used {\sl XSPEC~9.0} software (Arnaud et al. 1996) to fit a Raymond-Smith
plasma  model to the extracted spectra.
The temperature and abundance were allowed to vary in the global fit to the 
cluster.  
We also included a variable absorption component due to the column density of 
Galactic hydrogen in the line-of-sight.
The redshift of the model spectrum was fixed to 0.081, which corresponds to the 
heliocentric velocity of the cluster given by Burns et al. (1995). 
The extracted spectra were fit between $\sim$0.2 and 2.0 keV, the exact energy 
boundaries being set by the channel grouping. 
Once a minimum in $\chi^2$ was found, the 90\% confidence
errors were determined for the free parameters. 
A ``global" spectrum extracted from within a radius of 20\arcmin$\,$ from the
center yields a temperature of 3.71$^{+2.22}_{-0.14}$ keV and a (poorly
determined) metal abundance of less than 0.52 solar. 
This is in excellent agreement with the global temperature of $3.5\pm1.5$ keV 
found by Feretti et al (1997), who assumed an abundance of 0.35 solar.
The fitted value of N$_{\rm H}$ is 1.43$^{+0.18}_{-0.14}\times$
10$^{20}$ atoms cm$^{-2}$, less than the Galactic value of 
2.59$\times$10$^{20}$ atoms cm$^{-2}$ (Dickey \& Lockman 1990).

				\subsection{
Spatially Resolved Temperatures
				}
To examine the spatial distribution of temperatures, we extracted spectra
from a central core region and from 8 surrounding wedge-shaped regions 
(see Figure 1). 
The radius of the central region was chosen to enclose the area
where the (azimuthally averaged) cluster surface brightness profile shows an 
excess over the best-fitting analytic King model
(we segregated this ``excess" to optimize a search for the spectral signature 
of a cooling flow). 
This resulted in a circular region with radius 1\farcm 5 from the cluster
center.
The radii of the surrounding annular wedges were chosen so that each region 
has $\sim$ 10000 counts.

In fitting the spectra from the various subregions, we allowed the absorption
column to vary, but we fixed the abundance at 0.3 solar (since the abundance
was poorly determined in the global spectrum).  
The best-fitting temperature of the core region is 3.49$^{+3.67}_{-1.38}$ keV 
and the fitted N$_{\rm H}$ is $2.56^{+0.37}_{-0.40}\times$10$^{20}$ atoms cm$^{-2}$, 
both of which are consistent with the globally determined values.
There is no sign of a central temperature drop that would indicate the
presence of a cooling flow.  Cluster cooling flows also tend to be associated 
with centrally enhanced X-ray absorption, of which we see no evidence.
Our core temperature is consistent, within the errors, with the $RASS$ PSPC 
spectral analysis by Burns et al (1995), who found a core temperature 
of $1.9^{+2.3}_{-0.4}$ keV.
The temperatures for all of the individual sectors are shown in Figure 2 and 
Figure 3;
three of the nine regions (regions 2, 3 and 9) have 
temperatures that are significantly hotter than the mean cluster temperature
of 3.71$^{+2.22}_{-0.14}$ keV (which is indicated in Figure 3 by horizontal
lines).

				\subsection{
The Spectral ``Color'' Map
				}
To search for complicated temperature structure in the cluster which may have
been missed by using the regular regions in the spectral analysis above,
we generated a spectral ratio map of the cluster. 
Relatively soft and hard images were constructed from
0.52--1.31 keV photons (PSPC bands R5 + R6) 
and 1.13 -- 2.01 keV photons (band R7), respectively.
We used spectral simulations to find that the
ratio of these two bands offered the most dynamic range for a plausible
range of temperatures in Abell~2255.
After smoothing both images with a 1\arcmin$\,$ Gaussian,
the softer image was then divided by the harder image to create a ratio map. 
We used {\sl XSPEC} to determine the approximate temperatures corresponding to 
these ratios. 
Fixing the abundance to 0.3 solar and N$_{\rm H}$
to the best fit value found in the global spectral fit,
we find that these ratios are consistent with 
the temperatures found in the more detailed spectral analysis. 
This color map was generated in an attempt to trace the temperature 
structure of the cluster in more detail than was possible with the large 
regions used for the previous spectral analysis. However, because the 
global gas temperature (3.7 keV) is so far outside the $ROSAT$ 
bandpass (0.1--2.0 keV), 
any detail that this method might have revealed is lost in 
statistical noise. Only one relatively soft feature, northeast of the 
cluster center, stands out as statistically significant in the color map; 
we describe this soft feature below. 

Figure 4 shows the color ratios for an annular region, 
6\farcm 6--13\arcmin$\,$ about the cluster center, which has been divided 
into eight equal segments. All detected point sources were excluded from the 
calculation, using radii appropriate to the off-axis angles of the sources. 
A relatively soft (larger ratio) component is clearly seen in the northern 
portion of the annulus (0/360$^\circ$).  
The position of this soft region, at 
RA 17$^{\rm h}$13$^{\rm m}$15$^{\rm s}$ Dec +64$^\circ13^\prime$ (J2000),
is not associated with any 
optical object but is coincident with the diffuse radio ridge discovered
by Burns et al. (1995). The rough shape of the radio ridge is also seen in 
the color map and is indicated schematically as an ellipse in Figure 1. 
The value of the ratio in this region 
(5.00 $\pm$ 1.26) corresponds to a temperature of $\sim$ 1 keV, which is much 
cooler than the bulk of the cluster.  Using an elliptical aperture matched to 
the shape of the radio ridge we find that the the count rate from this region 
is 1.1$\times$10$^{-2}$ counts s$^{-1}$ which at the distance of the cluster 
corresponds to a luminosity of 3.6$\times$10$^{42}$ erg s$^{-1}$. 

We propose that this cool region is associated with an infalling group
of galaxies.  The temperature and luminosity of this cool region are consistent 
with those of the gas in a poor group of galaxies (Mulchaey et al. 1996).
The shocks associated with the entry of a group of galaxies into the
cluster can generate the relativistic electrons which power the radio ridge.

				\section{
Discussion and Conclusions
				}

Abell 2255 shows many signs of an ongoing major merger:
a large X-ray core radius, non-circular X-ray isophotes, complex
temperature structure, the lack of a cooling flow, a large galaxy 
velocity dispersion, a pair of central dominant galaxies, and a radio halo.
All of these signs are found in the Coma cluster, as well.
Simulations of cluster mergers show that strong shocks can heat
the gas to $\sim$10 keV (Evrard 1990; Roettiger et al. 1993;
Schindler \& M\"{u}ller 1993).
Our PSPC temperature map of Abell~2255 is too coarse to allow a detailed 
comparison with merger simulations, but it clearly indicates that there are
large, asymmetric temperature variations in the cluster gas.  The gas is 
particularly hot to the northwest of the core region.

We confirm previous work (Burns et al 1995; Feretti et al 1997)
which showed that the X-ray isophotes in the cluster core are also 
consistent with an ongoing merger: the isophotes are offset from the
central dominant galaxies and are fairly flat, with an ellipticity 
of $\sim$0.5; the isophotal ellipticity decreases outward until the 
outer regions are consistent with being round. A numerical simulation 
of this proposed merger shows that the X-ray isophotes should be elongated
in the direction of the merger (Burns et al. 1995). 

While there are many signs of a major merger in Abell~2255, we also find
X-ray evidence of a smaller ongoing merger event: $\sim13^\prime$ northeast
of the cluster center is a relatively cool ($\sim1$ keV) region, with
a luminosity comparable to a poor group of galaxies.
This cool region is coincident with a diffuse radio ridge discovered
by Burns et al (1995); the relativistic electrons which power this
radio source may themselves be powered by shocks driven by an infalling 
group of galaxies.  We suggest that a large sample of galaxy velocities 
(Ganguly, Hill \& Oegerle 1996) should be searched for kinematic
evidence of such an infalling group in the vicinity of
RA 17$^{\rm h}$13$^{\rm m}$15$^{\rm s}$ Dec +64\arcdeg 13\arcmin$\,$ (J2000).
Abell~2255 has evidently joined the ranks of clusters which are known to be 
undergoing major mergers, and like the Coma cluster, has a diffuse radio halo, 
a pair of dominate central galaxies and a cool group of infalling galaxies. 
In future work, we will use spatially resolved $ASCA$ spectra to 
prepare a more accurate temperature map for this cluster, to better
determine whether Abell~2255 is indeed a site of ongoing major and
minor subcluster merger events.

\acknowledgments
This research made use of the HEASARC, NED, and SkyView databases.
REW also acknowledges partial support from NASA grants 
NAG 5-1973 and NAG 5-2574.
DSD's research at M.I.T. is supported in part by the AXAF Science
Center as part of Smithsonian Astrophysical Observatory contract SVI--61010
under NASA Marshall Space Flight Center.

				\clearpage

\begin{table}
\caption{X-ray Data for Abell~2255}
\begin{tabular}{lcclr}
\tableline
\tableline
Region&Temperature&N$_{\rm h}$/10$^{20}$&$\chi^2$/dof\\
\tableline
      &keV&atoms cm$^{-2}$&\\
\tableline
Region 1&3.49$^{+3.67}_{-1.38}$&2.56$^{+0.37}_{-0.40}$&39.0/48\\
Region 2&11.84$^{+52.16}_{-7.14}$&1.89$^{+0.47}_{-0.42}$&55.9/45\\
Region 3&19.40$^{+44.60}_{-12.28}$&1.41$^{+0.42}_{-0.48}$&36.9/36\\
Region 4&2.66$^{+1.64}_{-0.82}$&2.65$^{+0.34}_{-0.38}$&44.3/46\\
Region 5&2.46$^{+2.12}_{-0.77}$&2.28$^{+0.62}_{-0.37}$&35.5/41\\
Region 6&3.71$^{+2.74}_{-1.23}$&0.85$^{+0.22}_{-0.23}$&78.3/82\\
Region 7&3.88$^{+5.16}_{-1.56}$&0.55$^{+0.28}_{-0.28}$&54.2/56\\
Region 8&4.75$^{+4.28}_{-1.69}$&1.32$^{+0.27}_{-0.29}$&106.3/92\\
Region 9&7.10$^{+11.07}_{-3.18}$&0.72$^{+0.23}_{-0.25}$&75.9/79\\
\tableline
\end{tabular}
\end{table}

				\clearpage
				\begin{figure}
				\title{
Figure Captions
				}
				\caption{
The regions for which the spectra are extracted for the 
PSPC derived temperature map for Abell~2255 are overlaying
the smoothed PSPC image of the cluster. The ellipse shows
the position of the radio ridge and the cool X-ray feature
found in the ratio map. 
				}
				\caption{
The best fit temperatures are shown on a region map like that seen
in figure 1. 
				}
				\caption{
The temperature for the regions shown in figure one are shown here.
The error bars are the 90\% confidence intervals. The vertical dotted lines
delimit the boundaries between the sectors seen in figure 1. The solid horizontal
lines show the average cluster temperature along with the upper and lower limits 
as dotted lines. 
				}
				\caption{
The color ratios for an annular region (6\farcm 5 -- 13\arcmin$\,$)
around the cluster center. A cool region (large ratio) can been seen in the 
two northern sections. 
				}

				\end{figure}


\clearpage
                                \begin{references}
                                \reference{}
Arnaud, K. A., Astronomical Data Analysis Software and Systems V, eds.
	Jacoby G. and Barnes J., p17, ASP Conf. Series 101.
                                \reference{}
Briel, U. G., Henry, J. P., Schwarz, R. A., Bohringer, H., Ebeling, H., 
	Edge, A. C., Hartner, G. D., Schindler, S., Trumper, J., Voges, W. 
	1991, A\&A, 246, L10
                                \reference{}
Burns, J. O., Roettiger, K., Pinkney, J., Perley, R. A., Owen, F. N. \& 
	Voges, W. 1995, ApJ, 446, 583
                                \reference{}
Bird, C. M. 1994, AJ, 107, 1637
                                \reference{}
Davis, D. S. \& Mushotzky, R. F. 1993, AJ, 105, 409
                                \reference{}
Dickey \& Lockman 1990, Ann. Rev. Ast. Astr., 28, 215
                                \reference{}
Edge, A. C., Stewart, \& Fabian, A. C. 1992, MNRAS, 258, 177
                                \reference{}
Evrard, A. E. 1990, in Clusters of Galaxies, Space Telescope Science Institute
	Symposium Series No. 4, edited by W. R. Oegerle, M. J. Fitchett, and 
	L. Danly (Cambridge University Press, Cambridge), p.287
                                \reference{}
Feretti, L., B\"{o}hringer, H., Giovannini, G., \& Neumann, D. 1997, \aap, 
	317, 432
                                \reference{}
Ganguly, R., Hill, J. M. \& Oegerle, W. R. 1996, BAAS, 188, 06.15
                                \reference{}
Hannish, R. 1982, A\&A, 116, 137
                                \reference{}
Henriksen, M. J. \& Markevitch, M. L. 1996, ApJ, 466, 79
                                \reference{}
Henry, J. P. \& Briel, U. G.1995, ApJ, 443, L9
                                \reference{}
Jaffe, W.J., \& Rudnick, L. 1979, ApJ, 233, 453
                                \reference{}
Jones, C. \& Forman, W. 1984, ApJ, 276, 35
                                \reference{}
Mulchaey, J. S., Davis, D. S., Mushotzky, R. F.\& Burstein, D. 1996, ApJ, 456, 80
                                \reference{}
Roettiger, K., Burns, J. O. \& Loken, C. 1993, \apj, 407, L53
                                \reference{}
Schindler, S. \& M\"{u}ller, E., 1993, A\&A, 273, 137
                                \reference{}
Snowden, S. L., McCammon, C., Burrows, D. N., \& Mendenhall, J. A. 1994, 
	ApJ, 424, 714
                                \reference{}
Stauffer, J., Spinrad, H. \& Sargent, W. L. 1979, ApJ, 228, 379
                                \reference{}
Stewart, G. C., Fabian, A. C., Jones, C. \& Forman, W. 1984, ApJ, 285, 1
                                \reference{}
Tribble, P. C. 1993, MNRAS, 263, 31
                                \reference{}
White, S. D. M., Briel, U. G. \& Henry, J. P. 1993, MNRAS, 261, L8
                                \reference{}
Zabludoff, A. I. \& Zaritzky 1995, Ap J, 447, L21

				\end{references}
				
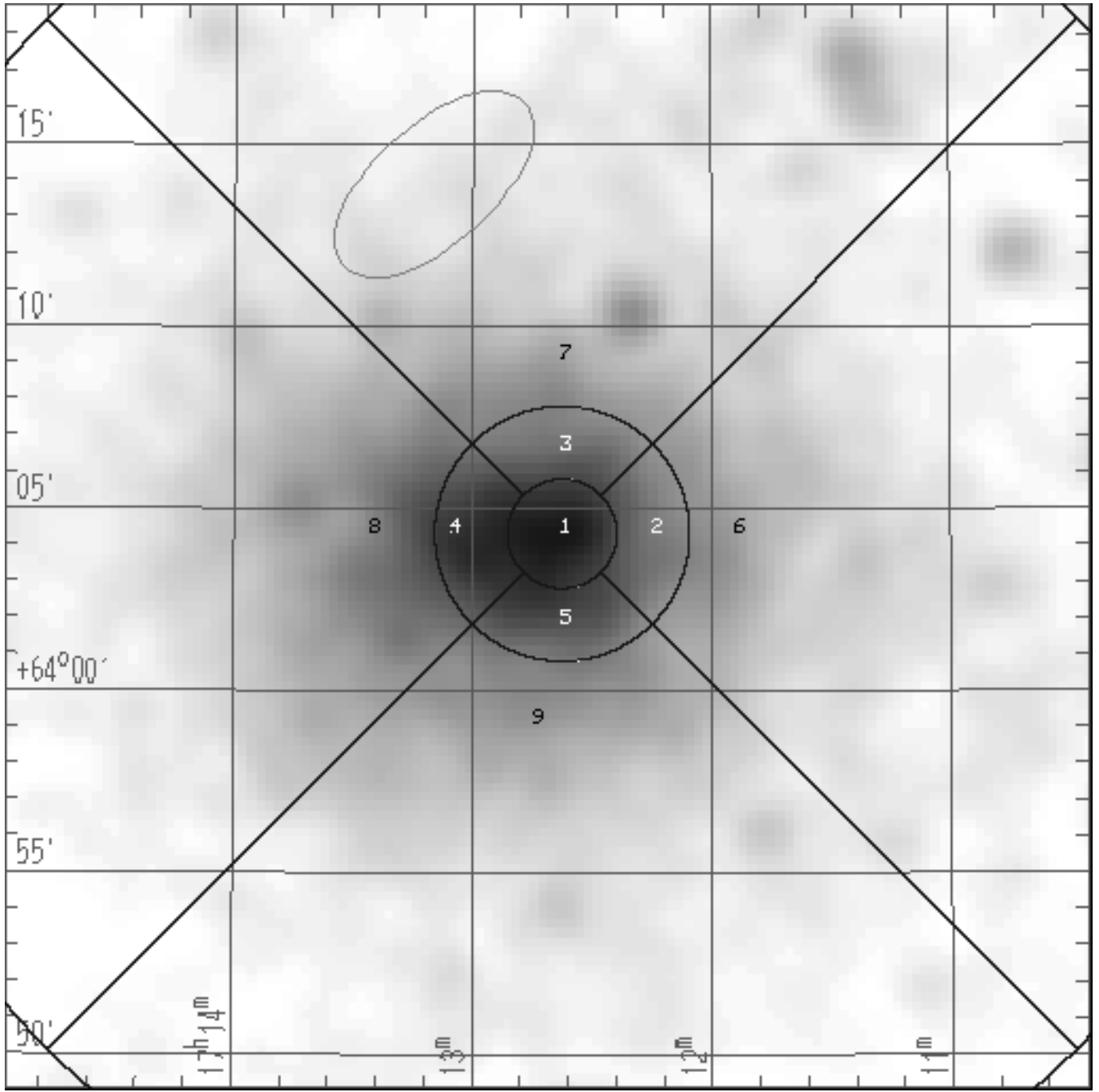
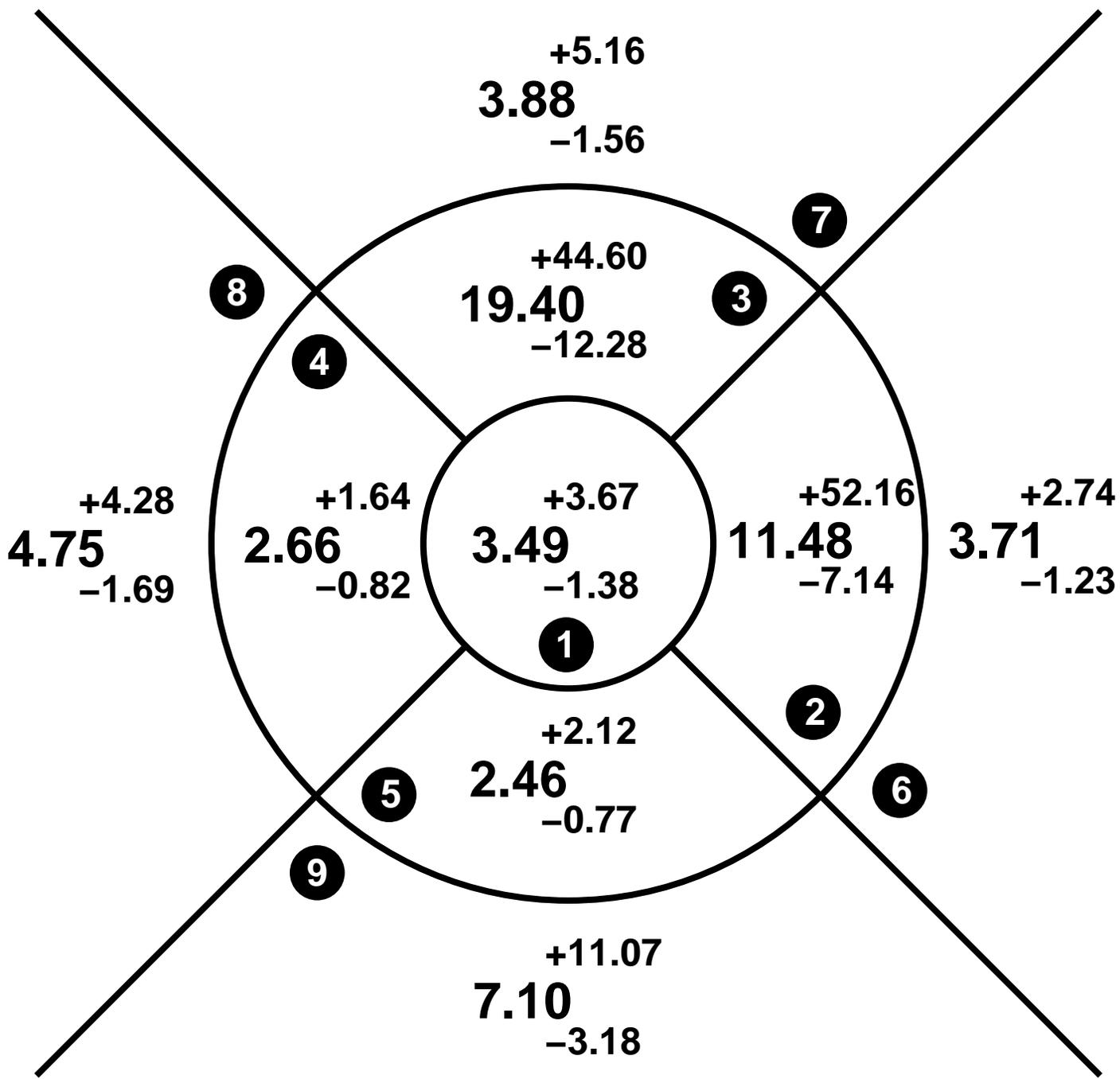
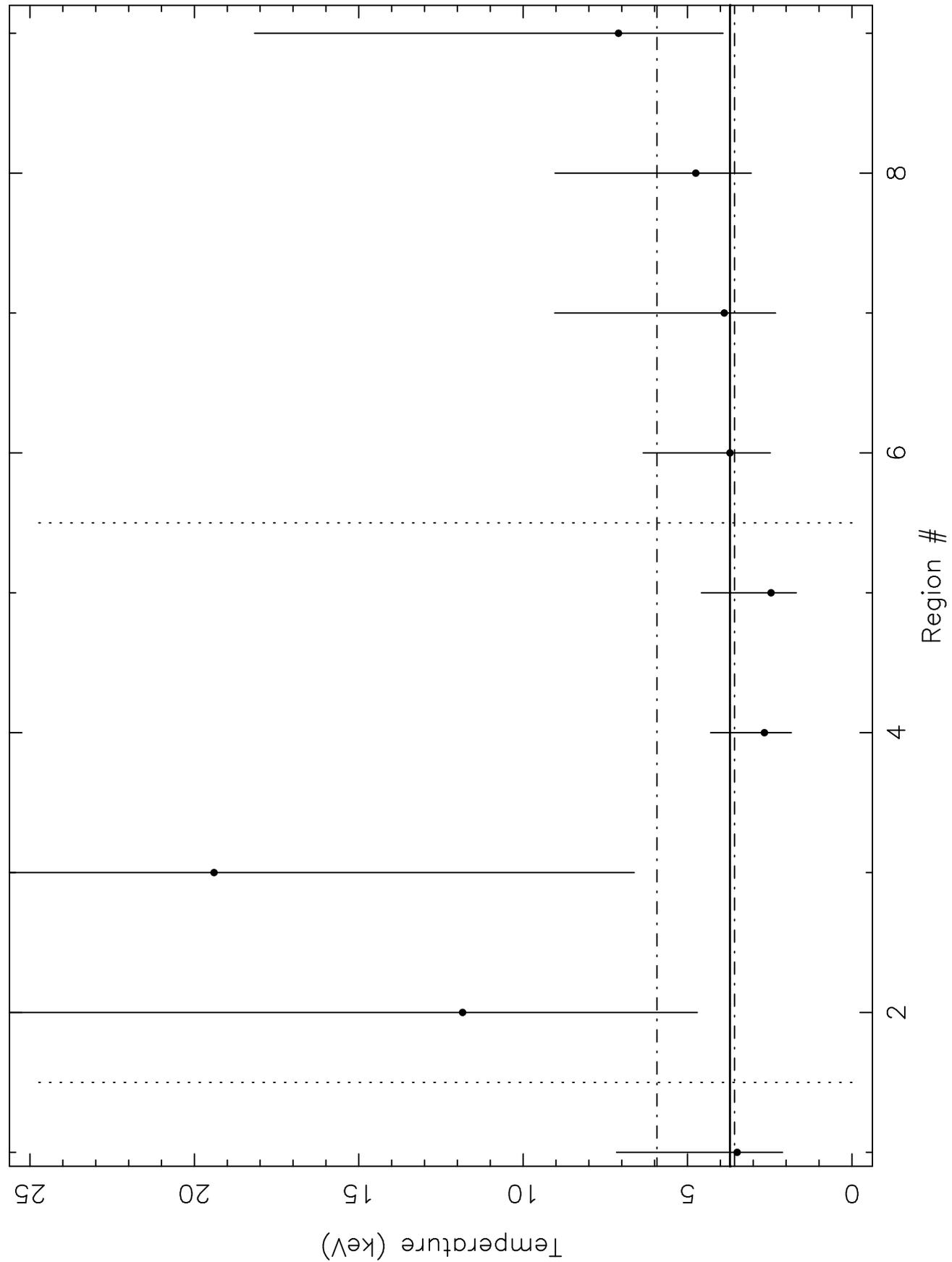
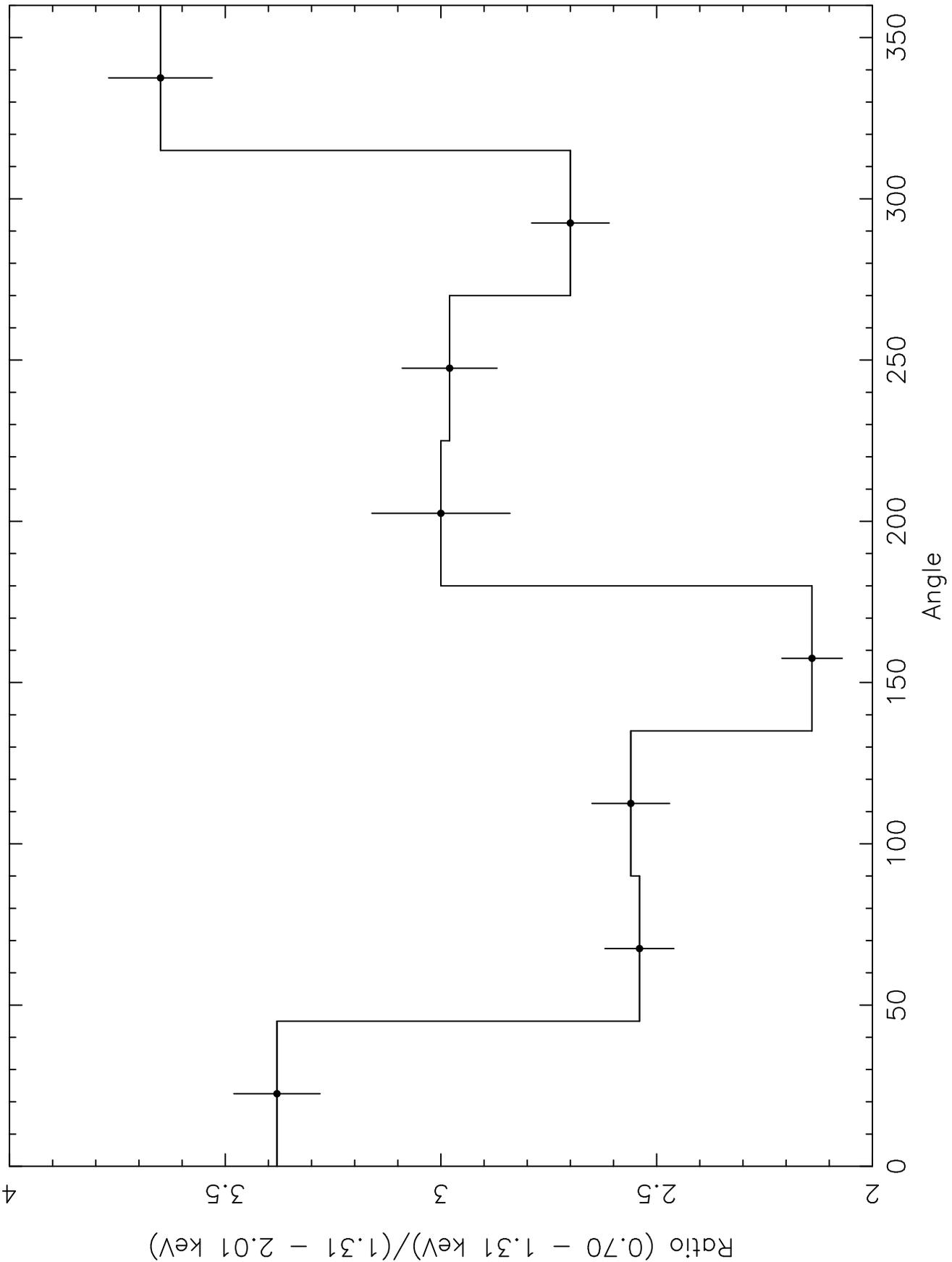
\end{document}